\documentstyle[12pt,aaspp4]{article}
\makeatletter
\def\lnyoro{\mathrel{\mathpalette\gl@align<}}
\def\gnyoro{\mathrel{\mathpalette\gl@align>}}
\def\gl@align#1#2{\lower.6ex\vbox{\baselineskip\z@skip\lineskip\z@\ialign{$\m@th
#1\hfil##\hfil$\crcr#2\crcr\sim\crcr}}}
\makeatother
\begin{document}
\title{\bf DUST-TO-GAS RATIO AND METALLICITY IN DWARF GALAXIES}
\author{\bf HIROYUKI HIRASHITA$^{1,2}$}
\affil
{$^1$:  Department of Astronomy, Faculty of Science, Kyoto University,
Sakyo-ku, Kyoto 606-8502, Japan}
\affil
{$^2$:  Research Fellow of the Japan Society for the Promotion of
Science}
\centerline{Feb.~3rd, 1999}
\centerline{email: hirasita@kusastro.kyoto-u.ac.jp}
\authoremail{hirasita@kusastro.kyoto-u.ac.jp}
\begin{abstract}

We examine dust-to-gas ratio as a function of metallicity for dwarf
galaxies [dwarf irregular galaxies (dIrrs) and blue compact dwarf
galaxies (BCDGs)]. Using a one-zone model and adopting the
instantaneous recycling approximation, we prepare a set of basic
equations which describes processes of
dust formation and destruction in a galaxy. Four terms
are included for the processes: dust formation from heavy elements
ejected by stellar mass loss, dust destruction in supernova remnants,
dust destruction in star-forming regions, and accretion of heavy
elements onto preexisting dust grains.
Solving the equations, we compare the result with observational data
of nearby dIrrs and BCDGs. The solution
is consistent with the data within the reasonable ranges of
model parameters constrained by the previous examinations. 
 This means that the model is successful in
understanding the dust amount of nearby galaxies. We also show that
the accretion rate of heavy element onto preexisting dust grains
is less effective than the condensation of heavy elements.

\end{abstract}


\section{INTRODUCTION}

Interstellar dust is composed of heavy elements made and ejected
by stars. Dwek \& Scalo (1980) demonstrated that
supernovae are the dominant source for the formation of dust
grains. They also showed that the dust is destroyed in supernova
shocks (see also McKee 1989, Jones et al. 1994, and
Jones, Tielens, \& Hollenbach 1996). Thus, the dust is formed and
destroyed in star-forming galaxies.

Some recent galaxy-evolution models treat the
evolution of total dust mass as well as that of metal abundance
(Wang 1991; 
Lisenfeld \& Ferrara 1998, hereafter LF98; Dwek
1998, hereafter D98; Hirashita 1999, hereafter H99).
The processes of dust formation and destruction by
supernovae are taken into account in LF98, in order to
explain the relation between dust-to-gas mass ratio and
metallicity of dwarf irregular galaxies (dIrrs) and blue
compact dwarf galaxies (BCDGs). We can find detailed mathematical
formulations to calculate the dust mass in any
galactic system in D98,
which treats the accretion of heavy elements onto preexisting dust
grains in molecular clouds in addition to the processes considered
in LF98.
The model by D98 is successfully applied to nearby spiral
galaxies in H99.

It is suggested in D98 that the accretion process onto
preexisting dust grains is not
effective in dwarf galaxies because of the absence of dense
molecular clouds there. If this is true, it is worth
estimating the ineffectiveness quantitatively, by which we can obtain
an information about molecular clouds in dwarf galaxies.
Direct observations of
molecular clouds in some dwarf galaxies have been extensively
carried out (e.g., Ohta, Sasaki, \& Sait\={o} 1988). However,
it is generally difficult to observe extragalactic
molecular clouds. Thus, theoretical constraints on the nature
of molecular clouds are indispensable for investigations on 
the star formation
processes in extragalactic objects.

In this {paper}, we examine the relation between dust-to-gas
ratio as a function of metallicity by using a set of equations
in H99. The model is one-zone (i.e., the spatial distribution
of physical quantities within a galaxy is not taken into account)
and the instantaneous recycling approximation is
applied. The main purpose of this paper is to apply the model
to dwarf galaxies (dIrrs and BCDGs).
First of all, in the next section,
we explain the model equation, through which the dust-to-gas ratio 
as a function of metallicity is
calculated. The result is compared with observational data of
dIrrs and BCDGs in \S 3.
We present a summary in the final section.

\section{REVIEW OF MODEL EQUATIONS}

In order to investigate the dust content in galaxies, H99
derived a set of equations describing the dust formation and
destruction processes, based on the models in LF98 and D98.
In H99, a simple one-zone model for a galaxy is adopted to
extract global properties of galaxies. For the model treating
the radial
distribution of gas, element, and dust in a galaxy, see D98.

The equation set is written as
\begin{eqnarray}
\frac{dM_{\rm g}}{dt} & = & -\psi +E-W,\label{basic1} \\
\frac{dM_i}{dt}       & = & -X_i\psi +E_i-X_iW, \\
\frac{dM_{{\rm d},i}}{dt} & = & f_{{\rm in},i}E_i-\alpha f_iX_i\psi
+\frac{M_{{\rm d},i}(1-f_i)}{\tau_{\rm acc}}-
\frac{M_{{\rm d},i}}{\tau_{\rm SN}}-\delta f_iX_iW.
\label{basic3}
\end{eqnarray}
(See eqs. [1]--[3] in McKee 1989, eqs. [6]--[8] in LF98 and
eqs. [1]--[3] in H99.)
Here, $M_{\rm g}$ is the mass of gas. The metal is labeled by
$i$ ($i={\rm O}$, C, Si, Mg, Fe, $\cdots$), and $M_{i}$ and
$M_{{\rm d},i}$ denote the total mass of the metal $i$ (in
gas and dust phases) and the mass of the metal $i$ in the dust
phase, respectively.  The star formation
rate is denoted by $\psi$; $E$ is the total injection rate of
mass from stars; $W$ is the net outflow rate from the galaxy;
$X_i$ is the mass fraction of the element $i$
(i.e., $X_i\equiv M_i/M_{\rm g}$); $E_i$ is the total
injection rate of element $i$ from stars; $f_i$ is the mass fraction
of the element $i$ locked up in dust (i.e., $f_i=M_{{\rm d},i}/M_i$). 
The meanings of the other parameters in the above equations are as
follows: $f_{{\rm in},i}$ is the value of the dust mass fraction in the
injected material, in other words, the dust condensation efficiency
in the ejecta\footnote{In this formalism, we assume that the
condensation efficiency in stellar winds is the same as that in
supernova ejecta.}; $\alpha$ refers to the efficiency of the dust
destruction during star formation [$\alpha =1$ corresponds to
destruction of
only the dust incorporated into the star, and $\alpha >1$ ($\alpha
<1$) corresponds to a net destruction (formation) in the star
formation]; $\tau_{\rm acc}$ is the accretion timescale of the
element $i$ onto preexisting dust particles in molecular clouds;
$\tau_{\rm SN}$ is the timescale of dust destruction
by supernova shocks; $\delta$ accounts for the dust content in
the outflow ($\delta =0$ means no dust in the outflow, while
$\delta =1$ indicates that the outflow is as dusty as the interstellar
medium).

We should comment on the parameter $\alpha$ here. Since the
protostellar disk forms dust, $\alpha <1$ is expected. However,
as to the circumstellar dust, the timescale of
loss of angular momentum through the Poynting-Robertson effect
is much shorter than the lifetime of stars (e.g., Rybicki \& Lightman
1979). This means that the formed dust is lost
effectively. Thus, we reasonably assume that $\alpha =1$ hereafter.
The formation of planets also contributes to the loss of the dust.

Here, we adopt the same assumption as LF98 and H99; the
instantaneous recycling approximation  (Tinsley
1980): Stars less massive than $m_{\rm l}$
live forever and the others die instantaneously.
This approximation allows us to write $E$ and $E_i$, respectively, as
\begin{eqnarray}
E & = & {\cal R}\psi , \\
E_i & = & ({\cal R}X_i+{\cal Y}_i)\psi ,
\end{eqnarray}
where ${\cal R}$ is the returned fraction of the mass that has
formed stars which is subsequently ejected into the interstellar
space, and ${\cal Y}_i$ is the mass fraction of the element $i$
newly produced and ejected by stars.\footnote{${\cal R}=R$ and
${\cal Y}_i=y(1-R)$ for the notation in LF98.}
${\cal R}$ and ${\cal Y}_i$ can be obtained using the following
formulae (Maeder 1992):
\begin{eqnarray}
{\cal R} & = & {\int_{m_{\rm l}}^{m_{\rm u}}(m-w_m)\phi(m)dm}
,\label{eqr} \\
{\cal Y}_i & = & {\int_{m_{\rm l}}^{m_{\rm u}}mp_i(m)\phi(m)dm}
,\label{eqy}
\end{eqnarray}
In equation (\ref{eqr}), $\phi (m)$ is the initial mass function
(IMF), and $m_{\rm u}$ is the upper mass cutoff of stellar mass.
The IMF is normalized so that the integral of $m\phi (m)$ in the
full range of the stellar mass becomes 1. Therefore, $\phi (m)$
has a dimension of the inverse square of the mass.
In equation (\ref{eqy}), $w_m$ is the remnant mass
($w_m=0.7M_\odot$ for $m<4M_\odot$ and $w_m=1.4M_\odot$ for
$m>4M_\odot$) and  $p_i(m)$ is the fraction of mass
converted into the element $i$ in a star of mass $m$.

Using the above parameters ${\cal R}$ and ${\cal Y}_i$,
and assuming that $W$ is proportional to the star formation rate
($W=w\psi$),
equations (\ref{basic1})--(\ref{basic3}) become
\begin{eqnarray}
\frac{1}{\psi}\frac{dM_{\rm g}}{dt} & = & -1+{\cal R}-w,
\label{basic4} \\
\frac{M_{\rm g}}{\psi}\frac{dX_i}{dt} & = & {\cal Y}_i,
\label{basic5} \\
\frac{M_{\rm g}}{\psi}\frac{d{\cal D}_i}{dt} & = & f_{{\rm in},i}
({\cal R}X_i+{\cal Y}_i)
-[\alpha -1+{\cal R}-\beta_{\rm acc}(1-f_i)+\beta_{\rm SN}-
w(1-\delta )]
{\cal D}_i,\label{basic6}
\end{eqnarray}
where ${\cal D}_i\equiv M_{{\rm d},i}/M_{\rm g}=f_iX_i$, and
$\beta_{\rm acc}$ and $\beta_{\rm SN}$ are, respectively,
defined by
\begin{eqnarray}
\beta_{\rm acc}\equiv\frac{M_{\rm g}}{\tau_{\rm acc}\psi}~~~
\mbox{and}~~~
\beta_{\rm SN}\equiv\frac{M_{\rm g}}{\tau_{\rm SN}\psi}.
\label{betas}
\end{eqnarray}
We can regard $\beta_{\rm acc}$ and $\beta_{\rm SN}$
as constant in time (\S 6.2 and \S 8.4 of D98).
We note that
the  Galactic value shows  $\beta_{\rm SN}\sim 5$ (LF98).
This value corresponds
to $\tau_{\rm SN}\sim 10^8$ yr, which is consistent with
Jones et al. (1994) and
Jones, Tielens, \& Hollenbach (1996).
The relation $\tau_{\rm acc}\simeq\tau_{\rm SN}/2$ (D98) leads to
$\beta_{\rm acc}\simeq 2\beta_{\rm SN}\simeq 10$.


Combining equations (\ref{basic5}) and (\ref{basic6}),
we obtain the following differential equation of ${\cal D}_i$
as a function of $X_i$:
\begin{eqnarray}
{\cal Y}_i\frac{d{\cal D}_i}{dX_i}= f_{{\rm in},i}
({\cal R}X_i+{\cal Y}_i)
-[\alpha -1+{\cal R}-\beta_{\rm acc}+\beta_{\rm SN}-w(1-\delta )]
{\cal D}_i-\frac{\beta_{\rm acc}{\cal D}_i^2}{X_i},
\label{difeq1}
\end{eqnarray}
where we used the relation, $f_i={\cal D}_i/X_i$.

Here, we take $i={\rm O}$ to compare the result with the data
in LF98.
For further quantification, we need to fix the values of
${\cal R}$ and ${\cal Y}_i$ for the traced element ($i={\rm O}$).
The values are given in LF98.
The reason why LF98 chose oxygen as the tracer is as follows:
(i) Most of oxygen is produced in Type II supernovae which are
also responsible for the shock destruction of dust grains; (ii)
oxygen is the main constituent of dust grains. The point
(i) means that the instantaneous recycling approximation may
be a reasonable approximation for the investigation of oxygen
abundances, since the generation of oxygen is a
massive-star-weighted phenomenon. In other words, results are
insensitive to the value of $m_{\rm l}$.
According to LF98, we put
$m_{\rm l}=1M_\odot$ and $m_{\rm u}=120M_\odot$.
We use a power-law form of the IMF: $\phi(m)\propto m^{-x}$.
The Salpeter IMF is assumed; i.e., $x=2.35$ (Salpeter 1955).
According to LF98, $({\cal R},\,{\cal Y}_{\rm O})=
(0.79,\, 1.8\times 10^{-2})$ 
for the Salpeter IMF. 
After the numerical integration of 
equation (\ref{difeq1}) by the
Runge-Kutta method, we compare the result with observational
data of dwarf galaxies in the next section. 

\section{APPLICATION TO DWARF GALAXIES}

To compare the solution of equation (\ref{difeq1}) with
observational data of dust-to-gas ratio, we make an assumption
that total mass of dust is proportional to that of oxygen
in the dust phase. In other words,
\begin{eqnarray}
{\cal D}\equiv\sum_i{\cal D}_i=C{\cal D}_{\rm O},\label{total}
\end{eqnarray}
where $C$ is assumed to be constant for all galaxies. According to
Table 2.2 of Whittet (1992), $C\simeq 2.2$ (the Galactic value).

We compare the solution of equation (\ref{difeq1})
with the data in LF98 (see also the references therein). The data sets
of nearby dIrrs and BCDGs
are presented in Tables 1 and 2 of LF98, respectively.
The observed dust-to-gas ratio is defined as
\begin{eqnarray}
{\cal D}^{\rm obs}\equiv M_{\rm d}/M_{\rm HI},
\end{eqnarray}
where $M_{\rm d}$ and $M_{\rm HI}$ are the total masses of dust
and H {\sc i} gas, respectively. The dust mass is derived from the
luminosity densities at the wavelengths of 60 $\mu$m and 100 $\mu$m
observed by {\it IRAS}.
The dust mass derived from the far-infrared emission is about an
order of magnitude smaller than the value found from the
analysis of the interstellar extinction (Fig.~2 of LF98). The
presence of cold or hot dust, emitting beyond 100 $\mu$m and
below 60 $\mu$m may be responsible for the discrepancy
(LF98). Thus, we should keep in mind that the
dust-to-gas ratio adopted here is underestimated in this
context. However, since we only take into account the H {\sc i}
gas as the gas content and do not consider H$_2$ gas, the
dust-to-gas ratio is overestimated (by a factor of $\sim 2$).
To sum up, we should keep in mind the uncertainty of an order of
magnitude for the dust-to-gas ratio derived from the observation
(${\cal D}^{\rm obs}$).

In the following subsections, we compare ${\cal D}$ calculated
by using equations (\ref{difeq1}) and (\ref{total}) with
${\cal D}^{\rm obs}$.
We focus on the two processes of dust formation: One is
the condensation of dust from heavy elements ejected by
stars, and the other is the accretion onto preexisting dust
grains. The efficiency of the former process is denoted
by $f_{\rm in,O}$ and that of the latter by $\beta_{\rm acc}$.
The latter process is not taken into account in LF98.
For the dependences of the relation between dust-to-gas ratio
and metallicity on 
IMF, see LF98 and H99.

\subsection{Dependence on $f_{\rm in,O}$}

We present the dependence of the result on the value of $f_{\rm in,O}$
in Figure 1a, in which
the solid, dotted, and dashed lines show the
${\cal D}$--$X_{\rm O}$ relation for $f_{\rm in,O}=0.1$, 0.05 and 0.01,
respectively. 
The other parameters are fixed to
$\alpha =1$, $\beta_{\rm acc}=2\beta_{\rm SN}=10$,
$\delta =1$ [i.e., $w(1-\delta )=0$]. Figure 1a shows that
the larger the efficiency of production of
dust from heavy elements is, the larger the dust-to-gas ratio
becomes. The data points represent the relations between
${\cal D}^{\rm obs}$ and $X_{\rm O}$ of dIrrs and BCDGs in LF98.
The filled and
open squares show the data points of the dIrrs and the BCDGs,
respectively.

In the limit of $X_{\rm O}\to 0$, the solution reduces to 
\begin{eqnarray}
{\cal D}_{\rm O}\simeq f_{\rm in,O}X_{\rm O}~~~\mbox{or}~~~{\cal D}
\simeq Cf_{\rm in,O}X_{\rm O}.\label{lowlim}
\end{eqnarray}
(See also LF98.) This means that ${\cal D}$ scales linearly with
$f_{\rm in,O}$ for the extremely low metallicity. Thus,
$f_{\rm in,O}$ can be
constrained by low-metal galaxies (see also H99).
Equation (\ref{lowlim}) means that we can constrain the parameter
$f_{\rm in,O}$ by examining 
the relation between ${\cal D}_{\rm O}/X_{\rm O}$
(the fraction of oxygen in the dust phase) and $X_{\rm O}$.
We show this relation in Figure
1b. The parameter adopted for each line is the same as Figure 1a.
We also show the data points of the same sample as Figure 1a,
assuming $C=2.2$ for all the galaxies to convert
${\cal D}^{\rm obs}$ into ${\cal D}_{\rm O}$.
Roughly speaking, our model predicts that
${\cal D}_{\rm O}/X_{\rm O}$ is constant in the range of 
the dwarf galaxies. This indicates that the low-metal
approximation used to derive
equation (\ref{lowlim}) is applicable to dwarf galaxies.
Thus, we can directly constrain the parameter $f_{\rm in,O}$
by dwarf galaxies. From the data points in Figure 1b, we see
$0.01\lnyoro {\cal D}_{\rm O}/X_{\rm O}\lnyoro 0.1$ or
$0.01\lnyoro f_{\rm in,O}\lnyoro 0.1$. We note that this range
is consistent with the analyses in H99.
However, we should keep in mind the uncertainty of the data
as described above in this section.

We also see from Figure 1b that the relation between
${\cal D}_{\rm O}$ and $X_{\rm O}$ becomes 
nonlinear  in the relatively high-metal region 
($\log X_{\rm O}> -3$). The behavior of this nonlinear
region depends on $\beta_{\rm acc}$ or $\beta_{\rm SN}$
(\S 3.2). 

\subsection{Dependence on $\beta_{\rm acc}$}

We here investigate the dependence of the solution on
$\beta_{\rm acc}$, which is proportional to the accretion efficiency
of heavy elements onto the preexisting dust grains (\S 2).
The resulting ${\cal D}$--$X_{\rm O}$ relations for various
$\beta_{\rm acc}$ are shown in Figure 2a. We show the cases of
$\beta_{\rm acc}=5,~10,~20$ (the solid, dotted, and dashed lines,
respectively), where the relation
$\beta_{\rm acc}=2\beta_{\rm SN}$ is fixed. The other parameters
are set to $f_{\rm in,O}=0.05$ ($\sim$ the center of the range
constrained in \S 3.1), $\alpha =1$, and $\delta =1$.
The value of
$\beta_{\rm acc}$ is determined by the lifetime of molecular clouds
(D98). The value $\beta_{\rm acc}\sim 10$ corresponds to
the accretion timescale of $\sim 10^8$ yr (\S 2).
The increase of $\beta_{\rm acc}$ means that the
accretion of heavy elements onto dust becomes efficient.
Thus, for a fixed value for the metallicity, dust-to-gas
ratio increases as $\beta_{\rm acc}$ increases.

We also present an extreme case of $\beta_{\rm acc}=0$ and
$\beta_{\rm SN}=5$ (long-dashed line). In this case, the accretion
onto preexisting dust
grains is neglected. We were not able to reproduce the
observational data of nearby spiral galaxies without
taking into account the accretion process
(D98; H99). However, the solid and long-dashed lines in Figure 2
show that we cannot judge whether
the accretion process is efficient or not because of the little
difference between the results with and without the accretion
process. 
Thus, the basic equations of LF98, which do not include the term
of the accretion were able to explain the observed relation between
dust-to-gas ratio and metallicity.
We note that the accretion process is
properly considered in D98.

The ineffectiveness of the accretion process
is understood as follows. Two processes are responsible for the
formation of dust in equation (\ref{basic3}); the dust condensation
from the heavy elements ejected by stars and the accretion of the
heavy elements  onto preexisting dust grains. The former is
expressed as $f_{{\rm in},i}E_i$ and the latter as
$M_{{\rm d},i}(1-f_i)/\tau_{\rm acc}$ (see eq. [\ref{basic3}]).
We define the following ratio, $A_i$:
\begin{eqnarray}
A_i\equiv\frac{M_{{\rm d},i}(1-f_i)/\tau_{\rm acc}}{f_{{\rm in},i}E_i}.
\end{eqnarray}
If $A_i< 1$, the accretion process is ineffective compared with the
dust condensation process. We will show that $A_i< 1$ for the sample
dwarf galaxies.

Using the instantaneous recycling approximation,
$A_i$ is estimated as
\begin{eqnarray}
A_i\simeq\frac{\beta_{\rm acc}(1-f_{i}){\cal D}_i}{f_{{\rm in},i}
({\cal R}X_i+{\cal Y}_i)}.
\end{eqnarray}
Now we put $i={\rm O}$. In Figure 2b, we present the relation
between $f_{\rm O}={\cal D}_{\rm O}/X_{\rm O}$ and $X_{\rm O}$.
The values of parameters for each line is the same as Figure 2a.
This figure shows that we can consider that $1-f_{\rm O}\sim 1$.
Moreover, in the range in which we are interested here,
${\cal D}_{\rm O}\simeq f_{\rm in,O}X_{\rm O}$ (eq. [\ref{lowlim}]),
and ${\cal R}X_{\rm O}\ll {\cal Y}_{\rm O}$ (as long as
$\log X_{\rm O}\lnyoro -2$ is satisfied).
Thus, $A_{\rm O}$ can be approximated by
\begin{eqnarray}
A_{\rm O}\simeq \frac{\beta_{\rm acc}X_{\rm O}}{{\cal Y}_{\rm O}}.
\end{eqnarray}
If we put $\beta_{\rm acc}=10$, 
and ${\cal Y}_{\rm O}=10^{-2}$, we obtain
$\log X_{\rm O}\lnyoro -3$ for the condition $A_{\rm O}<1$.
This is consistent with Figure 2b, since
the difference between the solid and the long-dashed lines
is clear for $\log X_{\rm O}> -3$. 
Thus, if $\log X_{\rm O}> -3$ is satisfied, the dust
accretion process is more ineffective than the dust condensation 
process. Actually, even for $\log X_{\rm O}\sim -2.5$,
the difference is within the typical error of the observed values.

The ineffectiveness of the accretion 
process results from the low metallicity. Thus, in galaxies with
high metallicity, the accretion process becomes important. Indeed,
D98 and H99 showed that the process is effective in spiral galaxies,
whose metallicity is much larger than the dwarf galaxies (H99).

Since we can reproduce the relation between dust-to-gas
ratio and metallicity of dwarf galaxies without considering the
accretion process, D98 suggested that the accretion process is not
efficient in dwarf galaxies because of the lack of dense molecular clouds.
This may be true, but
seeing that dIrrs and BCDGs show high
star formation efficiency (Sage et al. 1992; Israel, Bontekoe, \&
Kester 1996), there may be a large amount of
dense molecular gas in dwarf irregular galaxies.
Indeed, we have shown in Figure 2 that the observed relation can be
explained even if the efficiency of the accretion process
$\beta_{\rm acc}$ is as high as that in the spiral galaxies
considered in H99.

\section{SUMMARY AND DISCUSSIONS}

Based on the models proposed by LF98 and D98,
we have examined the dust content in dIrrs and BCDGs.
The basic equations which describes the changing rate of dust-to-gas
ratio include the terms of dust formation from heavy elements ejected
by stars, destruction by supernova shocks,
destruction in star-forming regions and
accretion of elements onto preexisting dust grains
(\S 2). This accretion
process is important in molecular clouds, where gas densities are
generally high. The results are compared with the observed values of
dIrrs and BCDGs. Though the degeneration of the
parameter and observational error makes it impossible to
determine each of the parameter precisely, we were able to
 constrain the parameters to some extent.

The efficiency of dust production  from heavy element
(denoted by $f_{{\rm in},i}$)
can be constrained by the galaxies with low metallicity (\S 3.1).
The reasonable range is  $0.01\lnyoro f_{{\rm in},i}\lnyoro 0.1$,
which is consistent with H99.
Thus, it is possible to understand the dust amount in dwarf systems
as well as that in spiral systems through the model in this paper.

As for the nearby spiral galaxies,
unless we take into account the accretion process of heavy
element onto the preexisting dust particles, 
we cannot explain the
observed relation between dust-to-gas ratio and metallicity
(D98; H99). For the dwarf galaxies,
however, we can explain the data without the accretion process
(\S 3.2). This means that the accretion
is not effective for dwarf galaxies.
Even if the efficiency of the accretion $\beta_{\rm acc}$, determined
by the lifetimes of molecular clouds (D98), is
as high as that in spiral galaxies, the accretion is not effective
because of the low metallicity in the dwarf galaxies.
Therefore, we cannot attribute the ineffectiveness of the dust
accretion process to the lack of molecular clouds.

Finally, we note that our model have satisfied one condition
which any model must fulfill: The model has to explain the
observation of {\it nearby} galaxies.
Then, it becomes a matter of concern whether our model can
explain the galaxies in the high-redshift Universe.
For theoretical modeling of the cosmic dust mass,
see, e.g., Edmunds \& Phillipps (1997).
Observationally, it is interesting that high-redshift
galaxies found recently have evidences of dust extinction
(Soifer et al. 1998; Armus et al. 1998).
The number count of galaxies in the far-infrared and submillimeter
wavelengths, where dust reprocesses stellar light, is
another interesting theme concerning high-redshift dust
(e.g., Takeuchi et al. 1999).

\acknowledgements
We would like to thank the anonymous referee for useful comments
which improved this paper.
We are grateful to S. Mineshige for continuous
encouragement. We thank H. Kamaya, K. Nakanishi,
T. T. Takeuchi and T. T. Ishii for kind helps and helpful comments.
This work was supported by the Research Fellowship of the Japan
Society for the
Promotion of Science for Young Scientists. We fully utilized the
NASA's Astrophysics Data System Abstract Service (ADS).

\newpage

\centerline{\bf FIGURE CAPTION}

\noindent
FIG. 1a---
The relation between the dust-to-gas ratio (${\cal D}_{\rm O}$) and
the oxygen abundance ($X_{\rm O}$)
for various $f_{\rm in, O}$ (the condensation efficiency of dust
from oxygen atoms).
The data points for dIrrs (filled squares) and BCDGs (open squares)
are from LF98.
The other parameters are fixed to $\beta_{\rm acc}=2\beta_{\rm SN}
=10$, $\alpha =1$, and $\delta =1$.
The solid, dotted, and dashed lines represent
different values of $f_{\rm in O}$ (0.1, 0.05, and 0.01, respectively).

\medskip

\noindent
FIG. 1b---
The relation between $f_{\rm O}={\cal D}_{\rm O}/X_{\rm O}$
(the fraction of oxygen in the dust phase) and
$X_{\rm O}$ (oxygen abundance). The values of the parameters and the
meanings of the data points are the
same as Fig.~1a.

\medskip

\noindent
FIG. 2a---
The same as Fig.~1a but for the different parameter sets
($f_{\rm in, O}=0.05$, $\alpha =1$, $\delta =1$
and various $\beta_{\rm acc}$ and $\beta_{\rm SN}$).
The solid, dotted, and dashed lines represent the cases
of $\beta_{\rm acc}=2\beta_{\rm SN}=5,~10,~20$, respectively.
The long-dashed line shows the case of no accretion process onto
the preexisting dust grains ($\beta_{\rm acc}=0$ and
$\beta_{\rm SN}=5$). The data points are identical to Fig.~1a.

\medskip

\noindent
FIG. 2b---
The same as Fig.~1b but for the parameter sets identical to
Fig.~2a. The data points are identical to Fig.~1b.

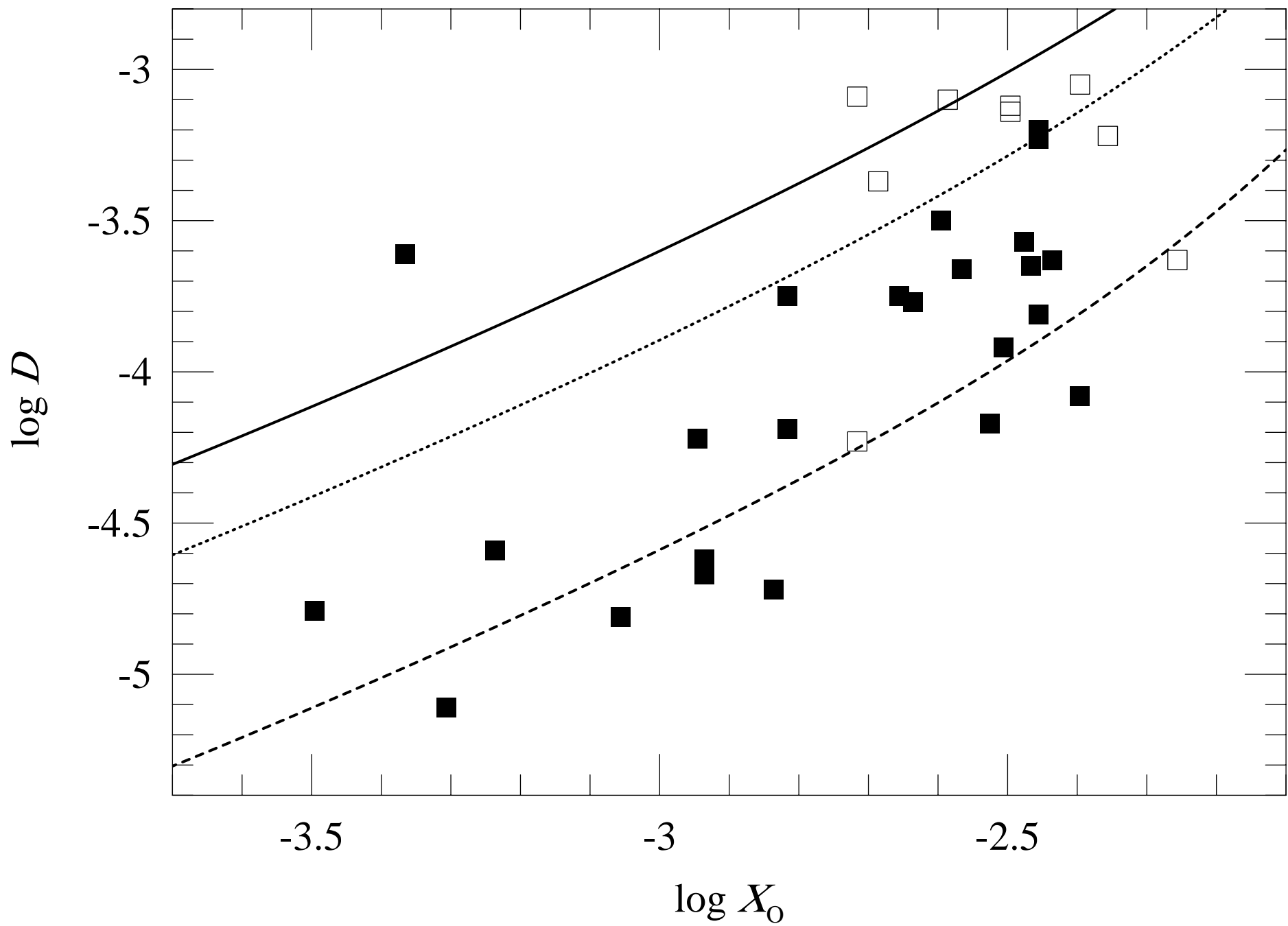
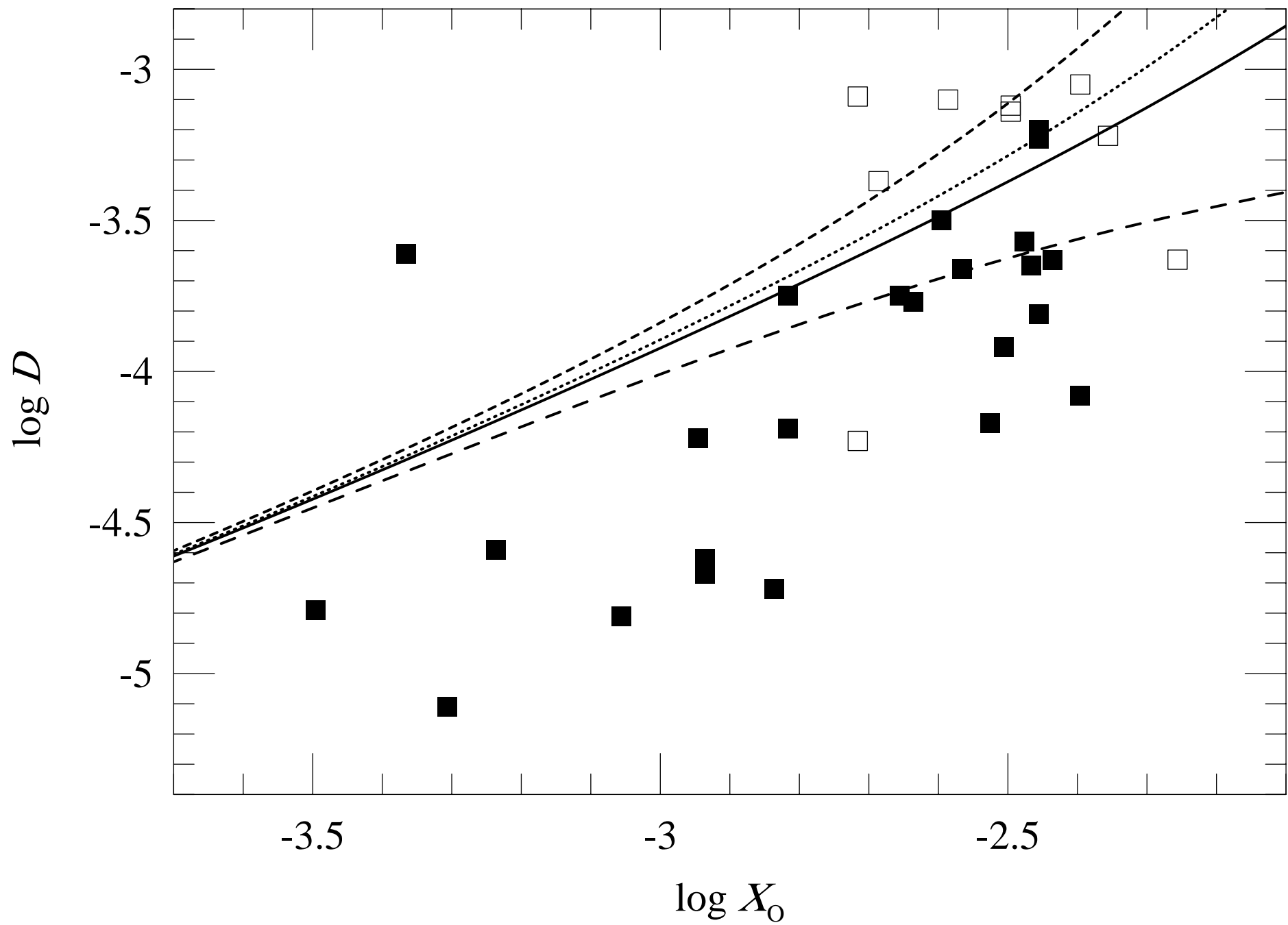
\end{document}